# Multi-Output Convolutional Neural Network for Improved Parameter Extraction in Time-Resolved Electrostatic Force Microscopy Data

*Madeleine D. Breshears,[a] Rajiv Giridharagopal,[a] David S. Ginger[a,\*]*

[a]Department of Chemistry, University of Washington, Box 351700, Seattle, Washington, 98195-1700, United States

[*]Corresponding Author: dginger@uw.edu

Time-resolved scanning probe microscopy methods, like time-resolved electrostatic force microscopy (trEFM), enable imaging of dynamic processes ranging from ion motion in batteries to electronic dynamics in microstructured thin film semiconductors for solar cells. Reconstructing the underlying physical dynamics from these techniques can be challenging due to the interplay of cantilever physics with the actual transient kinetics of interest in the resulting signal. Previously, quantitative trEFM used empirical calibration of the cantilever or feed-forward neural networks trained on simulated data to extract the physical dynamics of interest. Both these approaches are limited by interpreting the underlying signal as a single exponential function, which serves as an approximation but does not adequately reflect many realistic systems. Here, we present a multi-branched, multi-output convolutional neural network (CNN) that uses the trEFM signal in addition to the physical cantilever parameters as input. The trained CNN accurately extracts parameters describing both single-exponential and bi-exponential underlying functions, and more accurately reconstructs real experimental data in the presence of noise. This work demonstrates an application of physics-informed machine learning to complex signal processing tasks, enabling more efficient and accurate analysis of trEFM.

## 1. Introduction

Many systems, from new semiconductors for energy harvesting to new battery electrodes, exhibit electronic and ionic dynamics connected their nanoscale structure.[1–8] Probing these dynamics on natural length scales of the materials in question allows researchers to understand how morphological or structural heterogeneity influence performance.[4,6,9–13] Unlike optical microscopy, scanning probe microscopy (SPM) uses mechanical detection to map both morphological features and electronic properties of materials below the diffraction limit. These electronic properties include conductivity,[14,15] surface photovoltage,[3,5,6,16] and ion motion.[3,5,11] Time-resolved electrostatic force microscopy (trEFM) is a dynamic SPM technique that probes the evolution of the surface potential during photoexcitation in semiconductors, which is influenced by both electronic and ionic dynamics in mixed electronic/ionic conductors.[3,11,17] The perturbation of the surface potential is often assumed to follow a single-exponential response to an excitation, which is a satisfactory first approximation in many situations.[2,17–19] However, we also know that many systems are not well-characterized by a single time constant.[4,20] As a result, extending trEFM signal processing to account for multi-exponential dynamics would greatly extend its utility.

Machine learning (ML) tools have enabled efficient and accurate signal processing,[21–23] interpretation,[18] and reconstruction across a wide range of applications.[24–26] However, many traditional signal processing techniques rely on manual feature extraction[15,27] or domain-specific algorithms[22,28] that limit their ability to handle complex or noisy data. Recent advancements in deep learning, specifically convolutional neural networks (CNNs), have demonstrated potential for generalizable signal processing that is robust to noise.[22,24,28–30] Specifically, the application of CNNs to learn hierarchical patterns in complex signals poses an opportunity for efficient and accurate parameter estimation in time-series data,[21,24,30] such as signals collected via dynamic SPM methods like trEFM. In the trEFM signal, the transient carrier dynamics we are interested in measuring are intrinsically coupled to the cantilever physics. Previously, we used a single-exponential calibration procedure to extract cantilever-independent time constants to describe the dynamics of interest.[2,17,31] To improve the fit on noisy data, we developed a feedforward neural network that took the relevant cantilever parameters and the frequency signal and output a time constant corresponding to a single-exponential perturbation to the surface potential.[18] However, both of these methods relied on a single-exponential approximation of the underlying dynamics. Here, we present a multioutput convolutional neural network (CNN) trained on simulated and experimental data, and show this approach yields improved parameter extraction for both underlying single- and bi-exponential kinetics and improved robustness to noisy signals.

## 2. Methods

### 2.1 Perovskite sample preparation

All preparation of perovskite samples was performed in a $N_2$ glovebox. The 1.2 M $Cs_{0.17}FA_{0.83}Pb(I_{0.85}Br_{0.15})_3$ solution was prepared by mixing the correct molar ratio of $PbI_2$ (TCI), $PbBr_2$ (Alpha Aesar), FAI (Greatcell Solar), and CsI (Alpha Aesar) in a 4:1 solvent ratio of N,N-Dimethylformamide (DMF) and Dimethyl sulfoxide (DMSO). 1.5 cm$^2$ indium tin oxide (ITO) coated glass substrates were cleaned by sequential sonication in DI water containing ~2% Micro-90 detergent, DI water, acetone, and isopropanol for 10 minutes each. The substrates were ozone-cleaned for 23 minutes prior to spincoating. Me-4PACz (TCI) was dissolved in DMF at a concentration of 50 mg/mL and subsequently diluted to 1 mg/mL in isopropanol. Approximately 60 μL of the 1 mg/mL Me-4PACz solution was deposited on the substrate and spincoated at 5000 rpm for 30 s with an acceleration of 800 rpm/s. The film was annealed at 100°C for 10 minutes. Next, 100 μL of a solution with the ratio 1:150 $Al_2O_3$ nanoparticles (Sigma Aldrich) to isopropanol was spincoated at 6000 rpm for 30 s with an acceleration of 800 rpm/s to wash off excess Me-4PACz and improve wettability. The substrates were annealed at 100°C for 30 s. Finally, 70 μL of the perovskite precursor solution were deposited on the substrate and spincoated at 1000 rpm for 10 s with an acceleration of 200 rpm/s followed by spinning at 5000 rpm for 35 s with an acceleration of 800 rpm/s; 10 s before the completion of the final spincoating step, 300 μL of anisole was dynamically deposited on the sample. The sample was annealed at 100°C for 45 minutes.

*2.2 trEFM protocol*

trEFM and the required experimental setup have been discussed in detail.[17,31] We performed our measurements on an Asylum Research MFP3D-BIO mounted on a Nikon inverted optical microscope. We used Pt-coated cantilevers (mikroMasch HQ/NSC15/Pt) driven at resonance frequency (~300 kHz) for all measurements. We mounted the sample in an inert glovebox environment in a sealed cell, then imaged under active flowing nitrogen in this sealed cell. We recorded the cantilever oscillation using a 16-bit A/D digitizer (Dynamic Signals/GaGe Razor Express CSE1622), typically at 5 MS/s, and synced to the cantilever oscillation phase (180°) using custom trigger electronics (detailed circuit information can be found in our previous reports; additional information available upon request).[17,31] In an experimental window of 16 ms, we apply a bias of +10 V to the cantilever at t = 1 ms, we allow the sample to equilibrate for 4 ms. We digitize the cantilever oscillation starting at t = 4.6 ms through t = 8.6 ms, and trigger the laser excitation at t = 5 ms (turning off the laser at t = 7 ms). We used a 705 nm (Omicron PhoxXplus 705-40) continuous wavelength laser at an incident intensity of ~150 mW/$cm^2$ to excite our samples; we adjust the intensity using neutral density filters and the electrical power via software control. The laser was focused via a bottom objective on an inverted optical microscope and co-aligned with the cantilever tip. We measured the illumination intensity using the combination of a calibrated photodiode and a Pixera CCD camera (150CL-CU). We demodulated the raw cantilever deflection data using a Hilbert Transform to obtain the instantaneous frequency (Δω(t)). This code is freely available online via the FFTA Python package (https://github.com/GingerLabUW/FFTA).

*2.3 Multi-output Neural Network Approach*

We provide the code and example data online on GitHub (https://github.com/mdbresh/CNN_trEFM/). For the network itself, we used the PyTorch framework (v. 2.3.1 with Cuda 11.8). Our simulated training dataset varies the cantilever parameters k, Q, and ω according to the distributions in Supporting Information Fig. 1. We vary $τ_1$ between 1 and 10 μs, $τ_2$ between 50 and 500 μs, and A (the coefficient) between 0 and 1 according to the distributions shown in Supporting Information Fig. 2. During training, each label is normalized between 0 and 1 to prevent overcorrection during training and back propagation (generally, normalizing the labels makes the error gradients more uniform, leading to smoother and more efficient learning).[32–34] For initial training on simulated data, we use 10,000 traces total: 7,000 for training, 1,500 for validation, and 1,500 for testing. Results of random seed testing are shown in Supporting Information Fig. 3; mean average error of trained network and $R^2$ scores comparing CNN-extracted and true parameters are consistent across multiple random seeds, supporting the generalizability of the approach.

*2.4 Voltage Pulse Experiment to Obtain Labeled trEFM Data*

To obtain labeled experimental data for fine tuning, we use a function generator (Agilent 33500B) to apply voltage pulses that follow a bi-exponential decay to a conductive substrate and capture the cantilever response. The voltage pulse method is time-consuming, where each frequency traces takes approximately 1-2 minutes to obtain. We use 500 voltage pulse traces (350 for training, 50 for validation, and 100 for testing) for fine tuning.[35–37]

**3. Results and Discussion**

*3.1 The Limitations of the Single-exponential Model*

Equation 1 shows the bi-exponential function, which approximates realistic underlying kinetics in various systems with more than a single underlying physical process, such as surface potential equilibration in mixed ionic/electronic conductors[4] or semiconductors with two populations of traps.[20,38–41]

$$y = Ae^{-\frac{t}{\tau_1}} + (1-A)e^{-\frac{t}{\tau_2}} \quad \textbf{Equation 1}$$

Here A is the coefficient that describes the relative contribution of each time constant, $\tau_1$ and $\tau_2$; A is constrained to be between 0 and 1. When A = 0 or 1, the underlying function follows a single-exponential function. In practice, there will be an additional overall amplitude term to encode the strength of the perturbation; however, we ultimately normalize our signals between 0 and 1 to avoid convolving the amplitude with the time-dependent kinetics of interest.

We have discussed the working principles and analytical workflow of trEFM in previous publications (additional details are also provided in Methods 2.2).[4,17,31] Briefly, we drive an SPM cantilever at a known frequency; when we photoexcite a semiconducting sample, we generate electronic carriers and, depending on the material, induce ion motion.[4,11] The evolution of these charged carrier profiles results in a transient equilibration of the surface potential of the material.[4] The change in the surface potential perturbs the electrostatic force gradient between the oscillating tip and sample, resulting in a perturbation to the oscillation frequency of the cantilever ($\Delta\omega(t)$).[17,19,31] Because we are driving the cantilever at a fixed frequency, after the initial perturbation to the electrostatic force gradient the cantilever's instantaneous frequency relaxes back to the drive frequency, meaning the transient response we are interested in probing is both coupled to and obfuscated by the cantilever physics.

Fig. 1 shows the experimental response of a real cantilever to a bi-exponential perturbation generated with a programmed voltage pulse applied to a conductive substrate (see Methods 2.4). The $\Delta\omega(t)$ trace shows an initial fast shift to a point of maximum deviation from the drive frequency (normalized to 0), and then a slow relaxation back to equilibrium drive frequency (normalized to 1). The magnitude of the frequency shift encodes information about the relative strength of the perturbation, but not the time-dependent kinetics we are interested in probing.[31] Additionally, the magnitude may be dependent on the state of the cantilever's tip.[19,31] Previous work supports our hypothesis that the initial shift in Δω(t) corresponds to transient charge carrier dynamics of interest.[4,18] The subsequent relaxation back to the drive frequency is primarily governed by the cantilever Q.[17–19,31] For bi-exponential perturbations, we calculate the <τ> as a weighted average of the time constants using Equation 2.

$$<\tau> = A \times \tau_1 + (1 - A) \times \tau_2 \quad \textbf{Equation 2}$$

The dotted line in Fig. 1 shows a simulated signal governed by a single-exponential perturbation with a time constant of 170.12 μs, which we extracted using our previous feedforward neural network;[18] this τ markedly underestimates the true <τ> value of 277.85 μs. The simulated signal only approximately reconstructs our experimental data, with an $R^2$ score of 0.67, where a score of 1.0 would indicate perfect signal reconstruction. This degraded performance stands in contrast to the performance of the feedforward neural network on truly single-exponential data, extracting a τ value of 53.66 μs, compared to a true τ value of 58.81 μs, leading to a reconstruction $R^2$ score of 0.93.

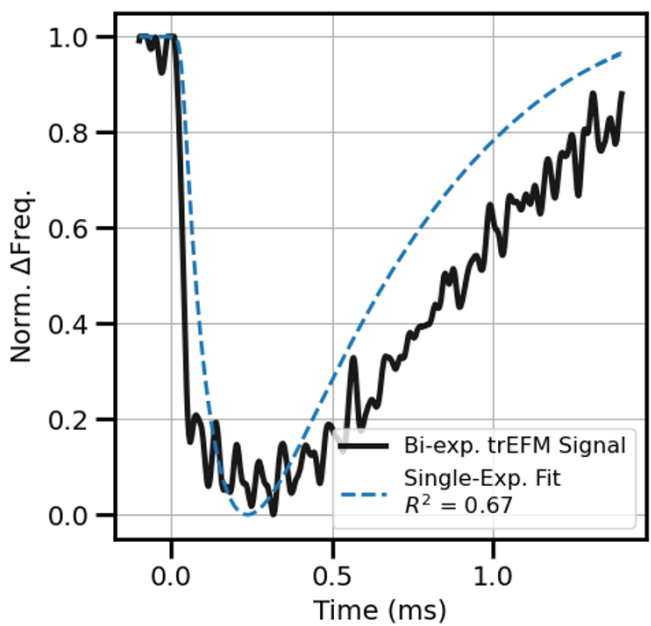

**Figure 1**: Example trEFM frequency trace governed by an underlying bi-exponential perturbation to the electrostatic force gradient with a <τ> of 277.85 μs, collected via the voltage pulse method described in Methods 2.4. The trigger for the bi-exponential voltage pulse is at t=0 ms. The Δω(t) trace is normalized such that a value of 1 represents the equilibrium drive frequency and a value of 0 represents the maximum shift in Δω(t). The blue dashed line is a simulated signal governed by a single-exponential perturbation with a τ value of 170.12 μs, extracted via a feedforward neural network from previous work.[18]

*3.2 The Multi-Output Convolutional Neural Network Model*

To regress to parameters that describe a more complex underlying function that follow Equation 1, we design a more complex neural network and train it on a broader set of simulated training data. We then test the trained network's performance on experimental data. Fig. 2 shows a schematic of our CNN architecture; the input includes the Δω(t) trace and the relevant physical cantilever parameters, which are easily obtained during the trEFM experiment. We use a PyTorch framework with a custom error function (see Supporting Information Note 1) and implement early stopping to prevent overfitting to our simulated data. We initially split the cantilever parameters from the input array containing the Δω(t) trace and we feed the Δω(t) signal into three convolutional branches. The convolution operation applies a filter (a kernel) to the signal, which helps the network detect and learn patterns in the data such as edges or noise.[28,42,43] We direct the network to pay attention to patterns of different timescales by changing the size of the kernel, where relatively smaller convolutional kernels learn faster patterns, like those in the initial shift in cantilever oscillation frequency, and relatively larger kernels learn patterns in the slower features, like the cantilever relaxation. In the fast and slow branches, we employ residual block layers to improve deep learning efficiency and gradient optimization for small variations in input data.[28,34,37] The weight branch is strictly a convolutional neural network, which we found best approximated the relative contributions from fast and slow processes.

When we train the network, each branch learns a latent representation of the Δω(t) trace. We concatenate the latent representations with the cantilever parameters and input those data into a shared features dense layer. In this step, we expect the network learns how the parameters contextualize the latent fast, slow, and weight patterns in the data. We direct the output of this dense layer into three branches that comprise two dense layers each. These simple feedforward branches regress on the contextualized latent representation of the signal and output each parameter: $\tau_1$, $\tau_2$, and A.

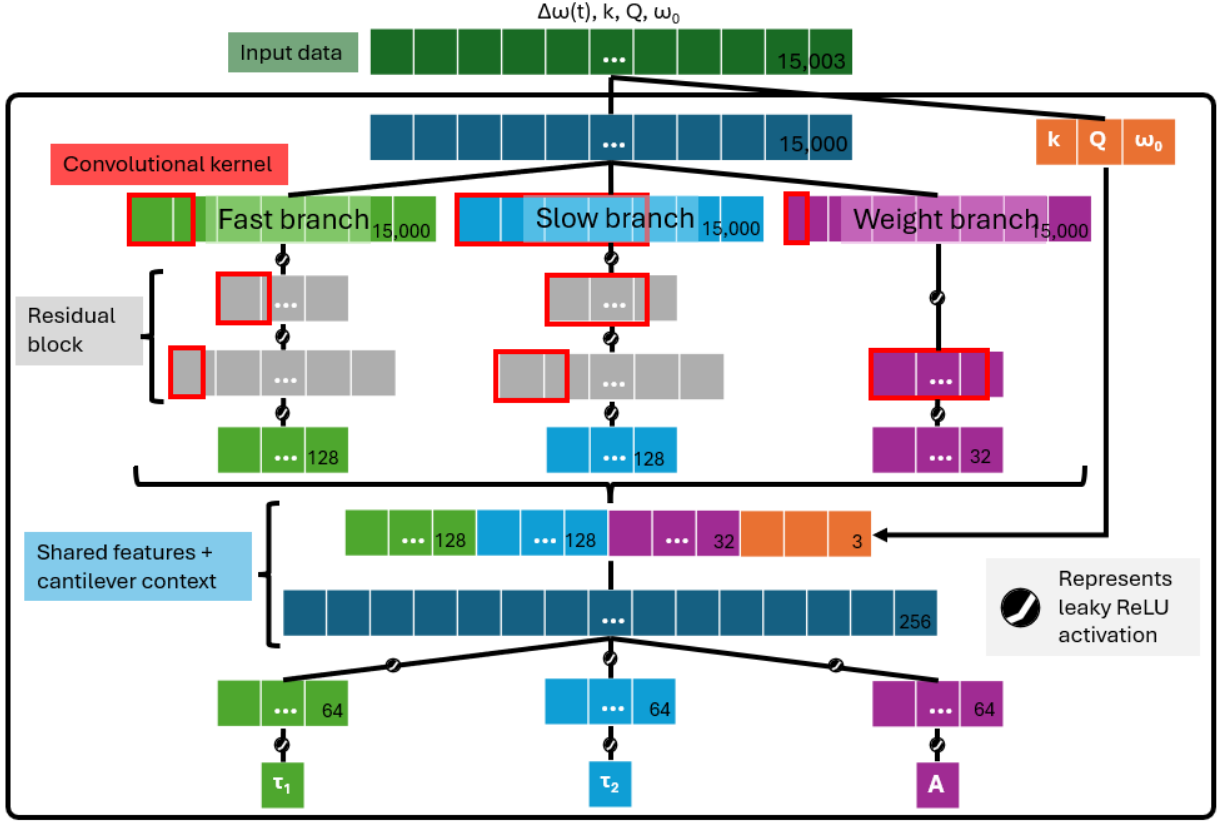

**Figure 2**: Schematic describing the multi-output convolutional neural network architecture, where the change in frequency ($\Delta\omega(t)$) is fed into three branches that learn the time-dependent features of the signal; the cantilever parameters (k, Q, and $\omega$) are then concatenated with these latent representations of the $\Delta\omega(t)$ signal. From the contextualized latent representations, three fully connected branches extract the relevant parameters given in Equation 1. Note that additional dropout regularization and batch normalization layers are not shown here; refer to https://github.com/mdbresh/CNN_trEFM/ for the full network architecture.

Our initial training dataset for our new network comprises simulated $\Delta\omega(t)$ traces and corresponding cantilever parameters. Supporting Information Fig. 1 shows the distributions of k, Q, and $\omega$ we use to simulate our training data. In our simulated training dataset, we vary $\tau_1$ between 1 and 10 μs, $\tau_2$ between 50 and 500 μs, and A between 0.0 and 1.0 (Supporting Information Fig. 2 shows distributions of these parameters used in our simulation). By constraining $\tau_1$ to relatively faster times and $\tau_2$ to slower times, the coefficient, A, represents the relative contributions of a "fast" and a "slow" component in the signal. From our previous work, we expect that the initial shift in frequency to its maximum deviation from the drive frequency contains the most information regarding the time constants, and that the following frequency relaxation is primarily governed by cantilever physics (specifically Q).[4,17–19,31] Supporting Information Fig. 4 shows how the initial transient response of the cantilever oscillation frequency is impacted by $\tau_1$, $\tau_2$, and A. After training the network, we evaluate its performance on simulated data; Supporting Information Fig. 5 shows the network's performance on simulated test data. We aim for the network to handle experimental noise sources, such as thermal noise, limitations from the photodiode, or jitter in the trigger.[44,45]

*3.3 Testing the Performance of the Trained CNN on Experimental Data*

To ensure the network learns to interpret realistically noisy data, we apply voltage pulses that follow single- or bi-exponential decays to a with known time constant parameters to a conductive substrate, providing us with a dataset of labeled experimentally obtained signals that captures real experimental noise sources. We obtain the cantilever parameters k (N/m), Q (unitless) and ω (kHz) during the experiment to include in our new dataset. We do not use solely the voltage pulse data to train our network, as collecting sufficient labeled data on multiple cantilevers would take too long; rather, we fine tune our CNN model that was already trained on simulated data with a smaller dataset of voltage pulse traces.[36] We discuss the voltage pulse method experimental details in Methods 2.4. Supporting Information Fig. 6 shows the parity plots of the test voltage pulse data, confirming that fine tuning enabled accurate parameter extraction on experimental data. The mean average error for $\tau_1$ was 1.01 μs, for $\tau_2$ was 35.87 μs, and for A was 0.04 (a mean percent error of 17.6%, 11.7%, and 8.3% respectively); and the $R^2$ score for each parameter's parity plot was 0.74, 0.79, and 0.97 respectively. Supporting Information Note 2 discusses the variations in error metrics for the three parameters in detail.

We believe this level of accuracy should be sufficient for many if not most imaging applications; the trained model separates the two time constants and their relative contribution to the underlying kinetics of interest. Because we cannot assign precise physical processes to each parameter, we use <τ> to summarize our network's overall performance. Fig. 3a shows a parity plot of <τ> extracted from voltage pulse data. The mean average error is 17.1 μs (~11.5% error), with an $R^2$ value (comparing CNN-extracted <τ> values to true <τ> values) of 0.97. Because trEFM is an imaging technique, we construct an artificial image using the labeled experimental test voltage pulse signals, where each pixel contains the Δω(t) trace from our voltage pulse dataset governed by a single- or bi-exponential underlying function. Supporting Information Fig. 7 shows the map of $\tau_1$, $\tau_2$, and A for this artificial image and the CNN-extracted parameter maps. Fig. 3b shows the <τ> map according to Equation 2. Fig. 3c shows the CNN-extracted <τ> for our artificial image.

We demonstrate that the trained multi-output CNN accurately extracts the parameters that describe a single- or bi-exponential underlying function; we expect that these parameters also enable us to more accurately reconstruct the Δω(t) signal. Given the CNN-extracted parameters, we simulate Δω(t) traces and calculate the $R^2$ value to compare the reconstructed signal with the experimental input. Supporting Information Fig. 8 shows the map of $R^2$ values for our artificial image, with an average $R^2$ of 0.97, confirming excellent signal reconstruction with the multi-output CNN. This performance is a significant improvement over the single-exponential feedforward network[18] which produces an average reconstruction $R^2$ score of only 0.79 (see Supporting Information Fig. 8).

Convolutional neural networks are a well-established method for extracting information (or fitting parameters) from noisy data.[24,28,33,46] We hypothesize that by using a multi-output CNN (that assumes bi-exponential underlying kinetics) we can not only extract parameters closer to the ground truth but also extract accurate parameters from noisy data. Supporting Information Fig. 9 shows the single-exponential feedforward network time constant and multi-output CNN <τ> values plotted against the signal-to-noise ratio (SNR) and the signal reconstruction $R^2$ score for each approach. From this analysis, we see that the multi-output CNN reconstructs the data with an $R^2$ score of 0.83 at an SNR as low as 5.85, whereas the single-exponential neural network results in a reconstruction $R^2$ score of 0.43. The multi-output CNN reconstruction $R^2$ score surpasses 0.9 when the SNR surpasses ~7, while the single-exponential neural network reconstruction $R^2$ score approaches 0.8 at an SNR of 15 (and never surpasses 0.81 even at higher SNR values). These results confirm that the multi-output CNN is more robust to noise than our previous feedforward network approach down to SNRs of ~7, indicating that the parameters extracted via the multi-output CNN more accurately represent the real dynamics of interest in the signal. Supporting Information Note 3 discusses the resilience of the CNN to Gaussian noise further.

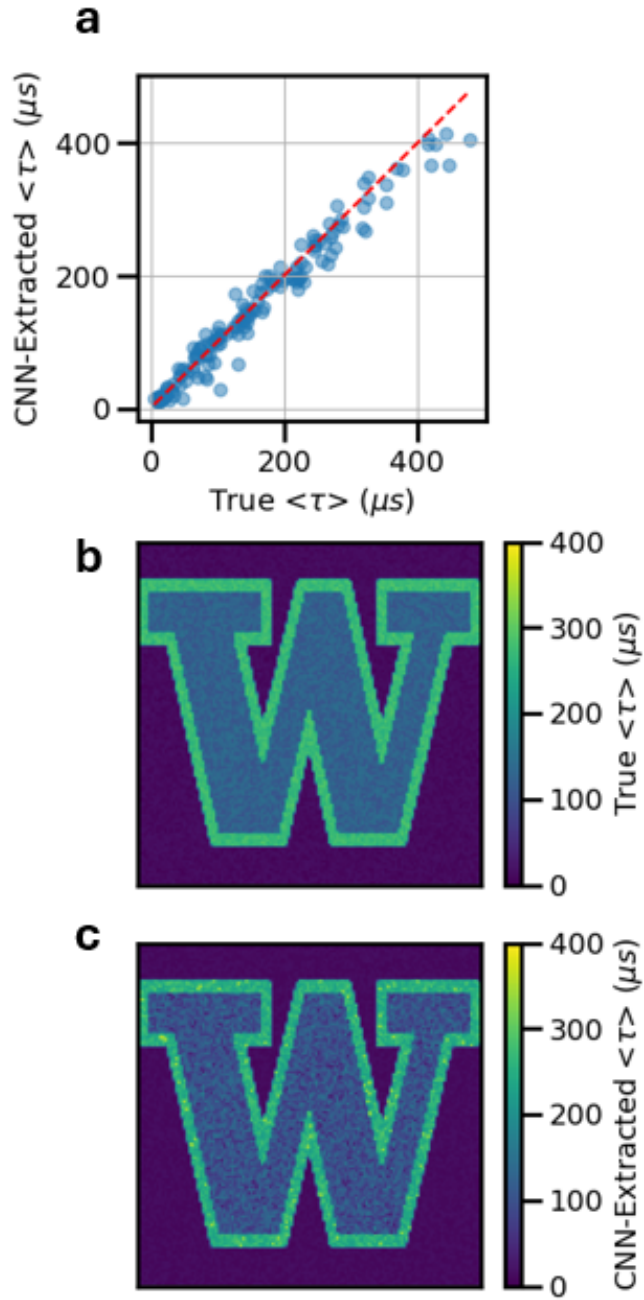

**Figure 3**: (a) Parity plot showing fine-tuned network performance on labeled experimental data; CNN-extracted <τ> values plotted against ground truth <τ> values calculated from Equation 2. (b) Artificially constructed ground truth image of <τ> values comprising labeled experimental data at each pixel. (c) CNN-extracted <τ> values recovered from the underlying experimental data at each pixel in the artificial image. Supporting Information Fig. 7 show individual parameter maps for ground truth and CNN-extracted $τ_1$, $τ_2$, and A; overall signal reconstruction $R^2$ value is 0.97 (where 1.0 would indicate perfect signal reconstruction).

### 3.4 Model Explainability

Given the success of our network in reconstructing the <τ> from labeled training data, we now turn to examining the patterns the model learned. Model explainability is critical to trust and transparency, bias detection, and user understanding of ML methods.[33,42,43,47] We previously used Shapley Additive exPlanations (SHAP)[47] to explore the impact of different features on our feedforward network's output, enabling us to confirm that the single-exponential network accurately interpreted the relevant cantilever parameters *and* learned intuitive patterns in the frequency signals.[18] Here, we apply SHAP to explore our more complex model and the impact of cantilever parameters and different parts of the frequency signal on each parameter output.

SHAP analysis reveals how every input into the network (e.g. Q for a particular cantilever or the Δω(t) value at a given index) indicates the model output should be greater than or less than (or equal to) the mean model output.[18,47] An input that the model learned corresponds to an output that is greater than the mean model output has a positive SHAP value,[47] and a negative SHAP value corresponds to an output less than the mean model output. If an input at a given index has no effect on the resulting model prediction relative to its mean output, the SHAP value will be 0.

We explore the distributions of SHAP values across the varied cantilever parameters in our simulated test dataset to confirm that the trained CNN has learned the appropriate cantilever physics. Supporting Information Fig. 9 shows the distributions of SHAP values across k, Q, and ω from this representative dataset. Importantly, Supporting Information Fig. 10 shows that the model learned that signals collected with cantilevers that have high Q values typically represent faster underlying dynamics, which not only physically intuitive[17,19] but also reproduces the results from our previous work.[18]

Fig. 4 shows three example Δω(t) traces colored by the SHAP value for the model output corresponding to the coefficient A. We see that the largest impact on the predicted coefficient, A, (the highest density of positive and negative SHAP values) lies along the initial frequency shift. The trace where A=0.1 has the slowest <τ> (per Equation 2) and exhibits the highest density of negative SHAP values along its transient frequency shift, indicating the model learned that a slow frequency shift results in a lower contribution of fast $τ_1$. In contrast, the trace where A=0.9 has the fastest <τ> (the largest contribution of fast $τ_1$) and exhibits a high density of positive SHAP values along its initial frequency shift, indicating the model learned that a fast frequency shift results in a higher contribution of fast $τ_1$. Supporting Information Figs. 11-12 show example traces for the SHAP analysis on $τ_1$ and $τ_2$. These results confirm that the initial transient response has the most impact on the model output, and that the slow cantilever relaxation has little to no effect on the model's output compared to the mean.

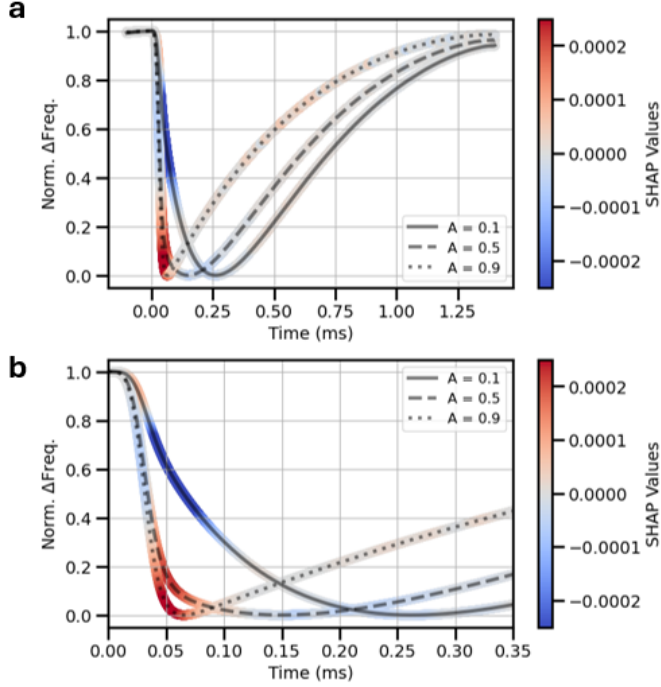

**Figure 4**: (a) Example Δω(t) traces with varied coefficient A values according to Equation 1, colored by SHAP values at each index, indicating the directional impact on the trained CNN each index has on the extracted coefficient A. Here, at t=0 ms the photoexcitation event is triggered. (b) Zoomed in on region immediately following trigger, showing varied densities of positive and negative SHAP values along initial transient frequency shift.

*3.5 Signal Reconstruction on Unlabeled Experimental Data*

Having demonstrated our multi-output CNN's ability to accurately reconstruct our data and withstand significant noise in the data and having explored feature importance with SHAP, we now apply the model to real, unlabeled experimental data. We performed trEFM on a mixed-cation, mixed-halide perovskite thin film device of the architecture and composition: ITO/Me-4PACz/$Cs_{0.17}FA_{0.83}Pb(I_{0.85}Br_{0.15})_3$. This composition is a useful, modern photovoltaic material for testing our approach on a real system of interest.[48] See Methods section for a detailed description of sample preparation and trEFM protocol.

After normalizing the Δω(t) trace at each pixel, we give the datacube to the trained CNN. Fig. 5a shows the 2×1 μm region of interest topography map. The sample exhibits 100-200 nm scale morphological features, commonly (if simplistically) referred to as "grains."[7,9,12] Given that these features cannot be resolved with optical microscopy, the fact that trEFM is a high-spatial resolution SPM technique makes it the ideal measurement tool for probing photoinduced dynamics.[4] Fig. 5b shows the <τ> map (according to Equation 2) we extracted with the multi-output CNN. We see that the grain boundaries exhibit slower surface potential equilibration dynamics than grain interiors, a phenomenon we have previously observed in these materials.[4,11] Supporting Information Fig. 13 shows the maps of $τ_1$, $τ_2$, and A, as extracted by the trained CNN. Given these parameters, we simulated Δω(t) traces for each pixel to evaluate how well the bi-exponential parameters reconstructed the experimental data. Supporting Information Fig. 14 shows the map of $R^2$ values comparing the experimental and simulated data, showing an average $R^2$ score of 0.89.

For comparison, we feed the same data into our single-exponential feedforward neural network; Supporting Information Fig. 15 shows the resulting τ map and the signal reconstruction $R^2$ map, with an average $R^2$ value of 0.71. These results indicate that the multi-output CNN extracts parameters that better reconstruct real experimental data, meaning the <τ> values we obtain with the CNN more accurately represent the charge carrier dynamics of interest.

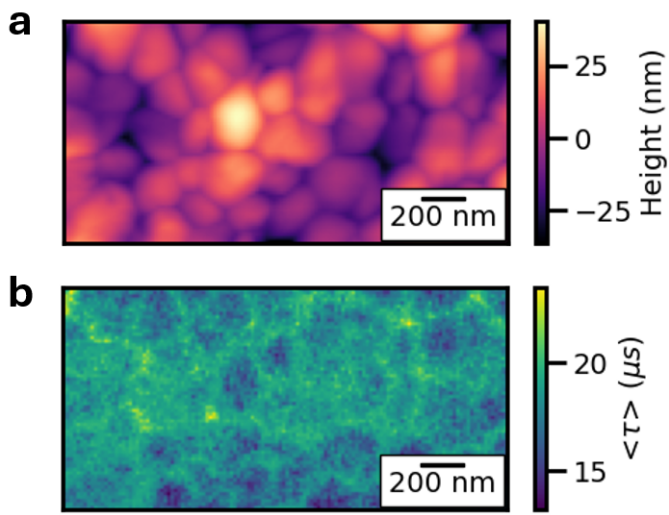

**Figure 5**: (a) Topography of 2×1 μm region of perovskite partial device (ITO/Me-4PACz/FA$_{0.83}$Cs$_{0.17}$Pb(I$_{0.85}$Br$_{0.15}$)$_3$), which exhibits 100-200 nm scale morphological features. (b) CNN-extracted <τ> values at each pixel. trEFM data were collected with a 705 nm laser at 150 mW/cm$^2$ incident intensity. Supporting Information Fig. 13 shows map of $R^2$ values comparing experimental data with simulated reconstructed signals, with an average $R^2$ value of 0.89.

### 4. Conclusion

We present a multi-branched, multi-output CNN trained on simulated data and fine tuned on labeled experimental data that efficiently and accurately extracts parameters that describe the underlying physical dynamics measured by trEFM. We demonstrate that the trained model extracts parameters which accurately reconstruct labeled and unlabeled experimental data. SHAP analysis reveals the importance of the initial shift in frequency on the extracted dynamics and confirms that the model correctly interpreted the role of cantilever physics on the resulting Δω(t) response. This work represents an application of physics-informed ML that offers a more robust and accurate analysis of dynamics in SPM signals, specifically of trEFM data, enabling more accurate signal reconstruction for interpreting the underlying kinetics of surface potential equilibration. The approach is highly generalizable to other underlying functions and techniques.

**Acknowledgement**

The atomic force microscopy imaging work was supported by the U.S. Department of Energy, Office of Basic Energy Sciences, Division of Materials Sciences and Engineering under Award DE-SC0013957. M.D.B. acknowledges support from the NSF Graduate Student Fellowship program under Grant no. DGE-2140004. D.S.G acknowledges salary and infrastructure support from the Washington Research Foundation, the Alvin L. and Verla R. Kwiram endowment, and the B. Seymour Rabinovitch Endowment.

**Data and Software Availability**

Example data, both simulated and experimental, and a guide for using the trained model can be found at https://github.com/mdbresh/CNN_trEFM/. The FFTA simulation package can be downloaded from https://github.com/GingerLabUW/FFTA.

Supporting Information for:

Multi-Output Convolutional Neural Network for Improved Parameter Extraction in Time-Resolved Electrostatic Force Microscopy Data

*Madeleine D. Breshears,[a] Rajiv Giridharagopal,[a] David S. Ginger[a],\**

[a]Department of Chemistry, University of Washington, Box 351700, Seattle, Washington, 98195-1700, United States

\*Corresponding Author: dginger@uw.edu

# Contents

**Supporting Information Fig. 1**: Distributions of cantilever parameters for simulated training dataset.

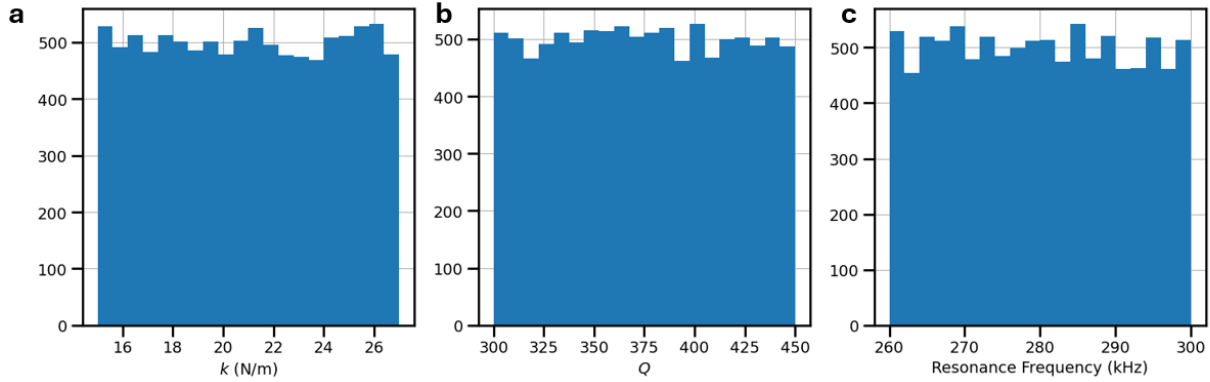

*(a) Distribution of cantilever spring constant, k (N/m), values used for simulated training data. (b) Distribution of cantilever quality factor, Q (unitless), values used for simulated training data. (c) Distribution of resonance frequency, $\omega_0$ (kHz), values used for simulated training data.*

**Supporting Information Fig 2**: Distributions of bi-exponential parameters used for simulated training datasets.

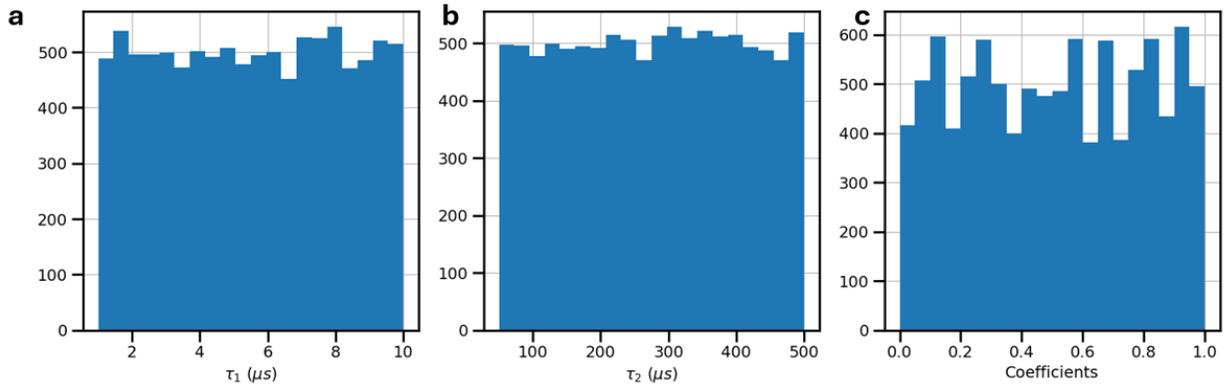

*(a) Distribution of $\tau_1$ (µs) values used in simulated training datasets. (b) Distribution of $\tau_2$ (µs) values used in simulated training datasets. (c) Distribution of coefficient, A, values used in simulated training. Refer to Equation 1 in the main text.*

**Supporting Information Fig 3**: Random seed testing shows approach is generalizable and not reliant on chosen seed.

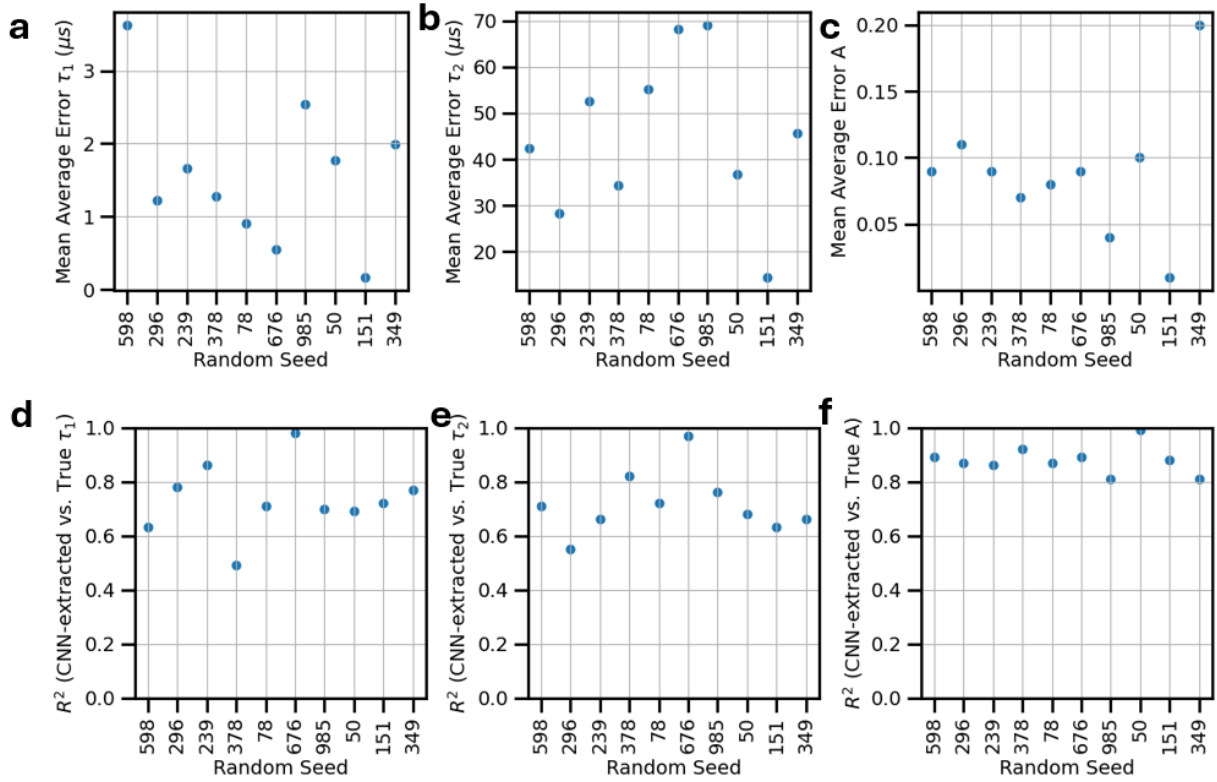

(a)-(c) Mean average error of trained network on $\tau_1$, $\tau_2$, and A respectively with varied random seed initializations, showing relatively consistent performance. (d)-(f) $R^2$ scores of CNN-extracted vs. true parameter values for $\tau_1$, $\tau_2$, and A. respectively, showing consistent, fair performance. Note that A is consistently predicted more accurately than each $\tau$ parameter, which we attribute to the normalization of the parameters during training leading to more accurate and consistent performance after un-normalizing.

**Supporting Information Fig 4**: Effect of bi-exponential parameters on cantilever frequency signal.

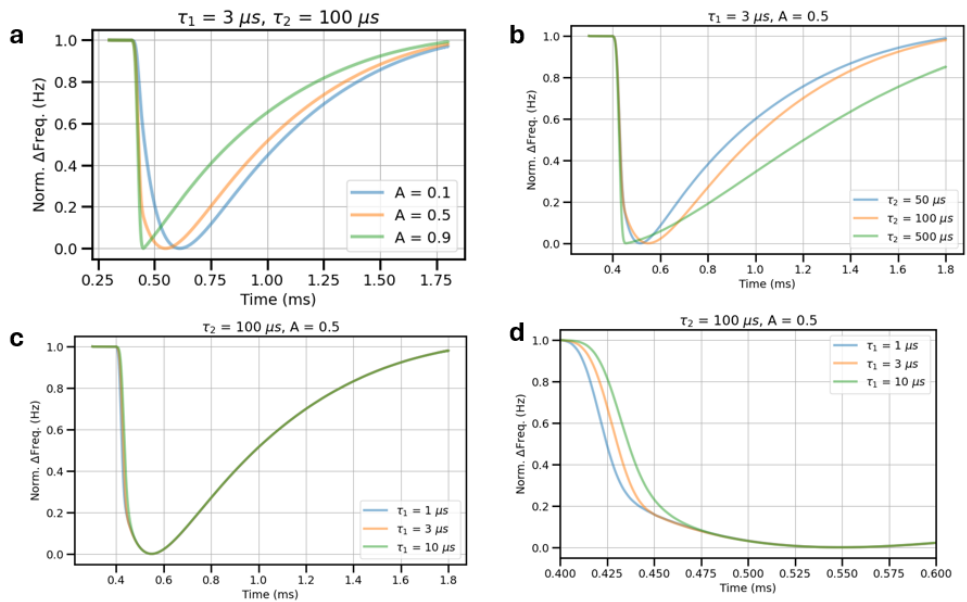

*(a) Example simulated Δω(t) traces following bi-exponential perturbations triggered at t=0.4 ms, where $\tau_1$ and $\tau_2$ are held constant at 3 μs and 100 μs respectively, and A (coefficient) is varied; refer to Equation 1 in the main text. (b) Simulated Δω(t) traces where $\tau_1$ and A are held constant at 3 μs and 0.5 respectively, and $\tau_2$ is varied. (c) and (d) show simulated Δω(t) traces where $\tau_2$ and A are held constant and $\tau_1$ is varied, showing slight variations in initial transient response after trigger at t=0.4 ms. All simulations performed with the FFTA code package (available at: https://github.com/rajgiriUW/ffta).*

**Supporting Information Note 1:** On the custom error function used in the multioutput CNN.

Given that each training label for $\tau_1$, $\tau_2$, and A (coefficient) is normalized between 0 and 1 to prevent exploding or vanishing gradients during training,[1,2] the error function used to penalize the network during training must reflect the difficulty of learning the patterns that correspond to each parameter. For instance, the subtleties of changing $\tau_1$ (while holding $\tau_2$ and A constant) as shown in Supporting Information Fig. 4 mean penalizing mistakes when predicting $\tau_1$ is more important than penalizing mistakes when predicting $\tau_2$ or A. After sweeping the coefficients for the mean squared error (MSE) of each parameter, we found the following combination enabled efficient and accurate learning of all three parameters:

$$criterion = 2.5 \times MSE(\tau_{1,true}, \tau_{1,pred}) + 1.75 \times MSE(\tau_{2,true}, \tau_{2,pred}) + 1 \times MSE(A_{true}, A_{pred})$$

Where MSE is the mean squared error between the true and predicted parameter.

**Supporting Information Fig 5**: Parity plots showing trained CNN performance on simulated test dataset.

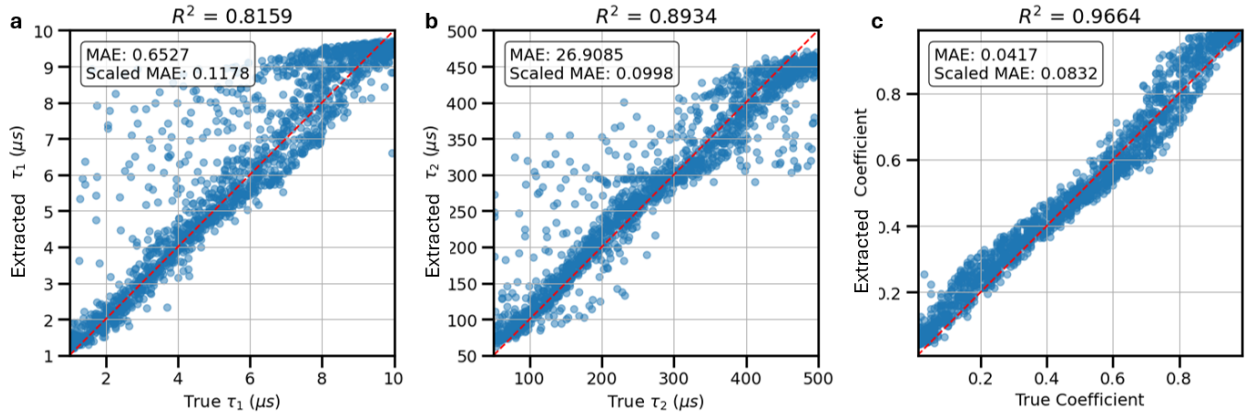

*(a) Parity plot showing trained (not fine-tuned) CNN performance on test simulated data when predicting τ₁ (μs). (b) Parity plot showing trained (not fine-tuned) CNN performance on test simulated data when predicting τ₂ (μs). (c) Parity plot showing trained (not fine-tuned) CNN performance on test simulated data when predicting coefficient, A. $R^2$ scores comparing the extracted value versus the true value are on the title of each plot. "MAE" stands for the mean average error; for τ₁ and τ₂ the MAE is in μs, and for A it is unitless. The "Scaled MAE" is the MAE divided by the average true value for the given parameter.*

**Supporting Information Fig 6**: Parity plots showing CNN performance on labeled experimental dataset after fine-tuning.

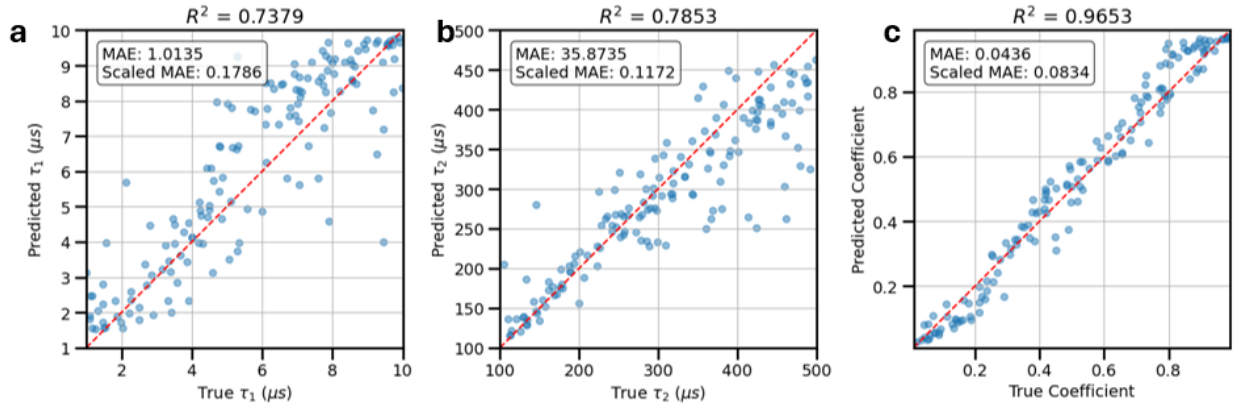

*(a) Parity plot showing fine-tuned CNN performance on test labeled experimental data when predicting τ₁ (μs). (b) Parity plot showing fine-tuned CNN performance on test labeled experimental data when predicting τ₂ (μs). (c) Parity plot showing fine-tuned CNN performance on test labeled experimental data when predicting coefficient, A. $R^2$ scores comparing the extracted value versus the true value are on the title of each plot. "MAE" stands for the mean average error; for τ₁ and τ₂ the MAE is in μs, and for A it is unitless. The "Scaled MAE" is the MAE divided by the average true value for the given parameter.*

**Supporting Information Note 2**: On the variation in error metrics for each parameter after fine-tuning.

We hypothesize that the network will extract the coefficient, A, with the most accuracy, because the relative contribution between fast and slow components shows a significant effect on the signal (Supporting Information Fig. 4). We also normalized all labels between 0 and 1 during training to prevent exploding gradients,[2] but A already ranged between 0 and 1, and as such we expect that the network will perform better when extracting this parameter. Not only are the effects of $\tau_1$ and $\tau_2$ on the $\Delta\omega(t)$ signal more subtle (Supporting Information Fig. 4), but by normalizing the labels corresponding to both $\tau_1$ and $\tau_2$, we expect that the resulting un-normalized error will be slightly larger compared to the error for A. The mean average error for $\tau_1$ was 1.01 μs, for $\tau_2$ was 35.87 μs, and for A was 0.04 (a mean percent error of 17.6%, 11.7%, and 8.3% respectively); and the $R^2$ score for each parameter's parity plot was 0.74, 0.79, and 0.97 respectively, as expected.

**Supporting Information Fig 7**: Artificial image parameter truths and extracted values.

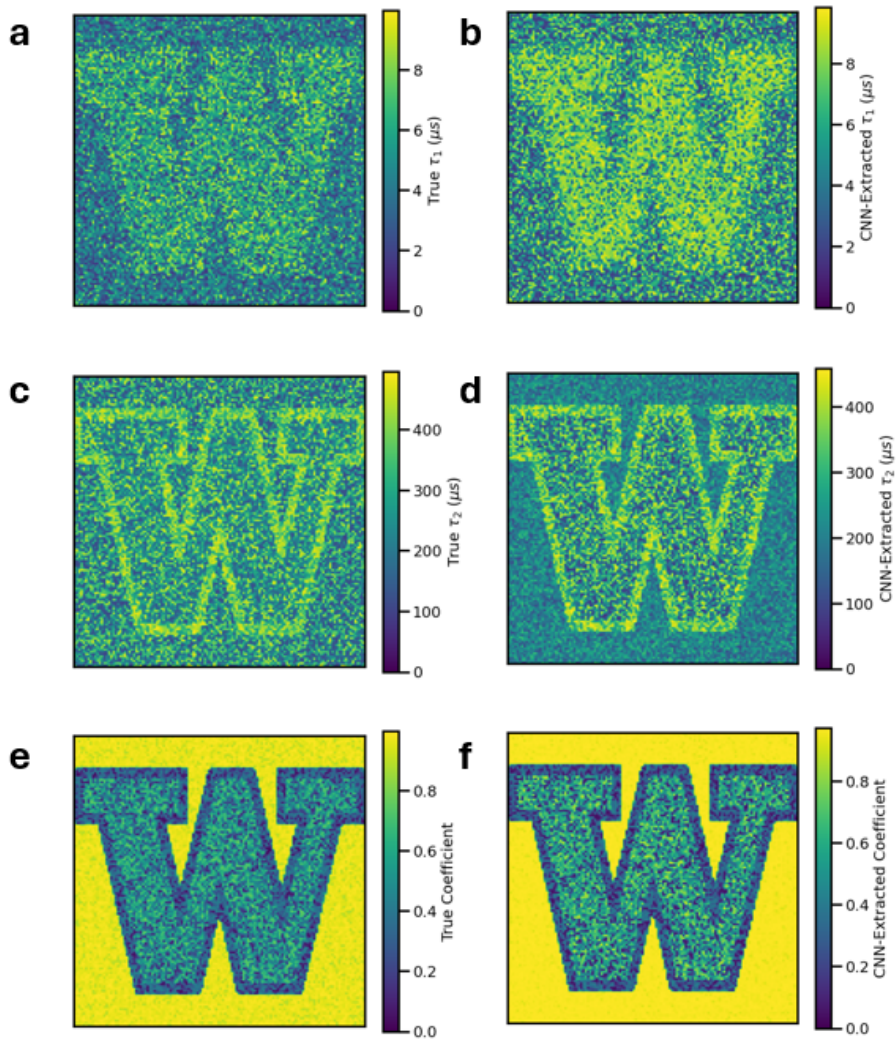

*(a), (c), and (e) show the map of each parameter ($\tau_1$, $\tau_2$, and A) that constructs the artificial image shown in main text Figure 3. (b), (d), and (f) show the CNN-extracted parameter.*

**Supporting Information Fig 8**: Comparison of fine-tuned CNN and single-exponential neural network on artificial image.

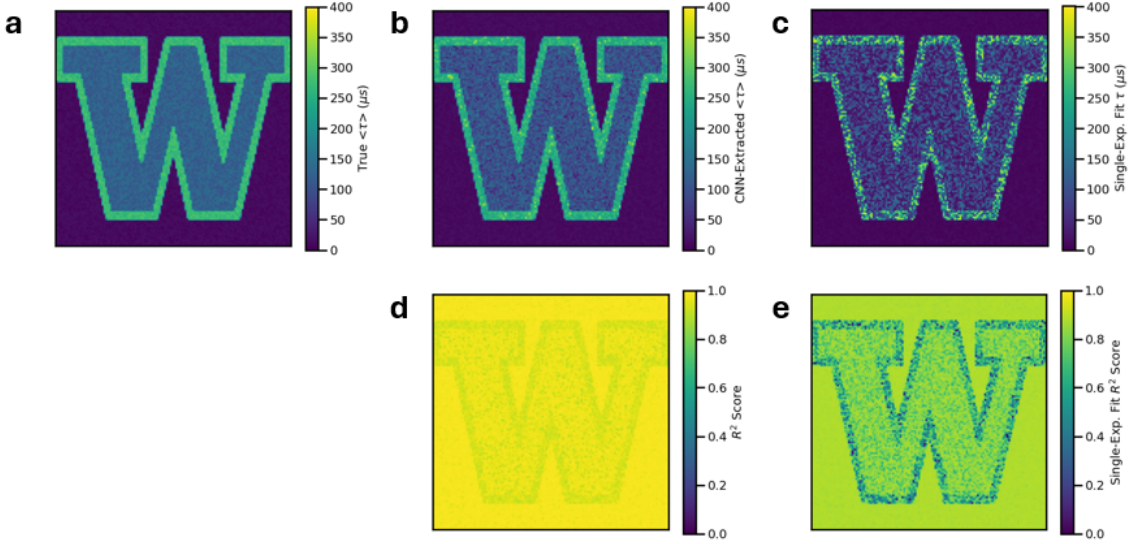

(a) Artificially constructed <τ> image (see Equation 2 in main text). (b) CNN-extracted <τ> image, showing relatively good agreement with (a) by eye. (c) Map of τ values obtained from feeding artificial image (with bi-exponential perturbations) into single-exponential feedforward neural network from previous work,[3] showing increased noise. (d) Map of $R^2$ values comparing the true data used to construct the artificial image compared to Δω(t) traces simulated from CNN-extracted parameters, where an $R^2$ value of 1 would indicate perfect signal reconstruction. (e) Map of $R^2$ values comparing the true data used to construct the artificial image compared to Δω(t) traces simulated from the single-exponential fit shown in (c).

**Supporting Information Note 3:** On adding artificial Gaussian noise to simulated instantaneous frequency signals.

We collect the cantilever's raw oscillation and then demodulate it to its instantaneous frequency. To accurately mimic this process, we introduce Gaussian noise to a simulated cantilever oscillation before demodulating it to obtain the instantaneous frequency. We add the noise according to the following instructions:

1. $\overline{P_{signal}} = mean(Z^2)$, where $P_{signal}$ is the average power of the signal Z.
2. We convert that to decibels via: $\overline{P_{signal,dB}} = 10 * \log(\overline{P_{signal}})$.
3. We calculate the power of the desired noise level (in decibels) with: $P_{noise,dB} = P_{signal,dB} - SNR_{desired}$.
4. $\overline{P_{noise}} = 10^{\frac{P_{noise,dB}}{10}}$, allows us to calculate the overall power of the noise.

We generate an array of Gaussian noise centered at 0 with a standard deviation of the square root of $\overline{P_{noise}}$ to add to our clean simulated signal (Z).

After we demodulate the signal (see main text Methods), we are left with the transform of that Gaussian noise, meaning our SNR$_{desired}$ value does not accurately represent the noise level of the signal. To obtain SNR$_{final}$, we do the following:

1. $noise = Z_{noisy} - Z$ gives us the noise by subtracting the clean signal (Z) from the noisy signal (Z$_{noisy}$).
2. $\overline{P_{signal}} = mean(Z^2)$ gives us the power of the signal, just like before.
3. $\overline{P_{noise}} = mean(noise^2)$ gives us the power of the noise.
4. To calculate the SNR$_{final}$, we need to compare the powers in decibels. We can obtain that ratio according to the following: $SNR_{final} = 10 * \log\left(\frac{\overline{P_{signal}}}{\overline{P_{noise}}}\right)$.

Supporting Information Figs. 9a-c show example simulated Δω(t) traces with artificial Gaussian noise applied according to this method. Supporting Information Figs. 9d-e show how well the multi-output CNN extracts accurate <τ> values from increasing noisy (decreasing SNR) simulated data. (Supporting Information Figs. 9d-e also compare the single-exponential neural network's performance.) At SNRs of as low as ~7 the multi-output CNN parameters enable data reconstruction with an R² score of 0.9, meaning the dynamics we extract assuming a bi-exponential perturbation more accurately represent the dynamics we are interested in at higher noise levels.

**Supporting Information Fig 9** Evaluation of the limits of noise on single-exponential feedforward neural network and multi-output convolutional neural network.

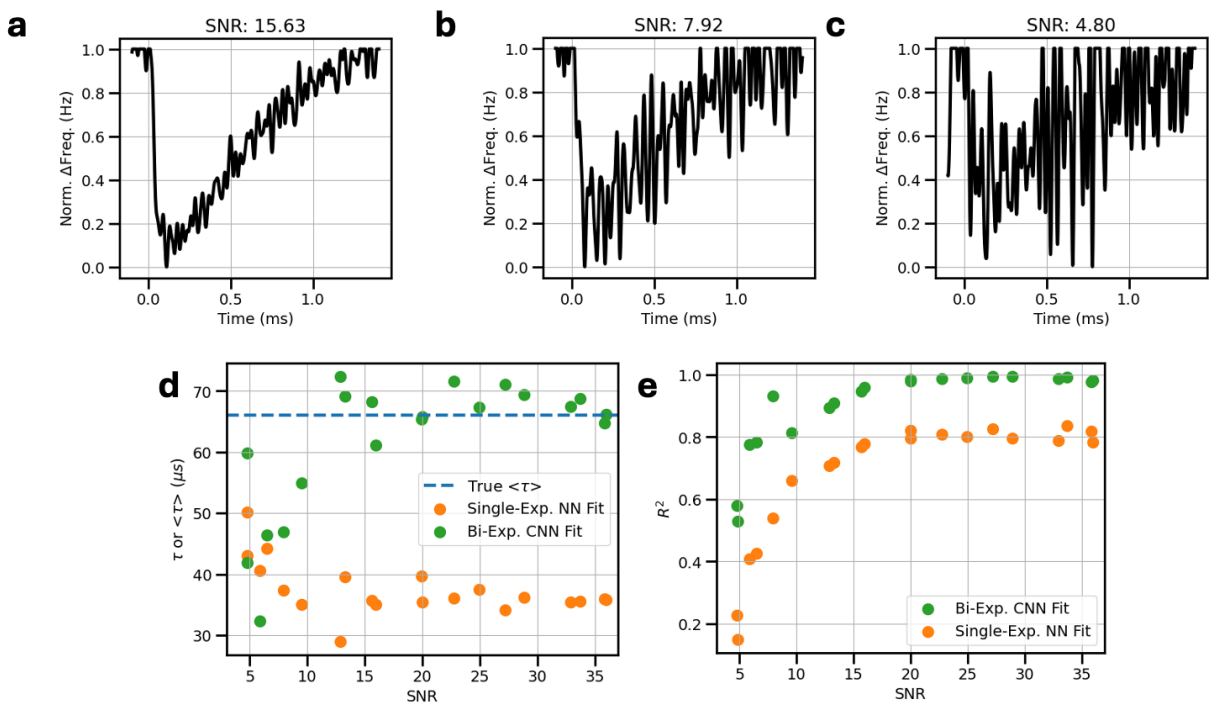

(a)-(c) Example simulated Δω(t) traces with varied levels of Gaussian noise added via the process described in Supporting Information Note 3. (d) Scatter plot showing single-exponential neural network[3] and multioutput CNN performance on a simulated signal with increasing levels of Gaussian noise. Refer to Supporting Information Note 3 for discussion on how noise was added to the simulated signal. (e) R² score comparing how well each network (single-exponential or multioutput CNN) reconstructed noisy data according to increasing signal-to-noise ratio, showing that multioutput CNN better reconstructs noisy data.

**Supporting Information Fig 10**: SHAP analysis exploring the impact of each cantilever parameter.

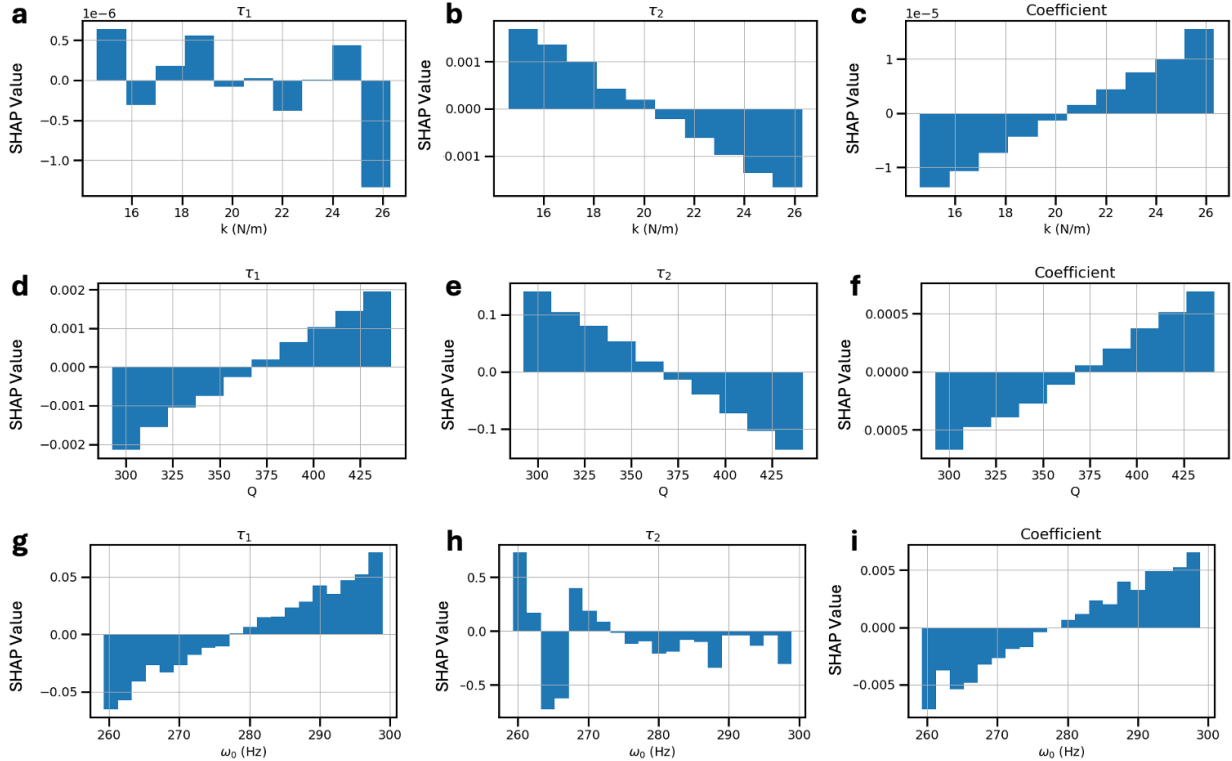

(a)-(c) SHAP value distributions plotted against cantilever spring constants (k, N/m) showing how varied spring constants affect the overall model output for each parameter ($\tau_1$, $\tau_2$, and A); note how small the SHAP values are, indicating that k has little to no effect on the model output. (d)-(f) SHAP value distributions plotted against cantilever quality factors (Q, unitless) showing how varied quality factors affect the overall model output for each parameter ($\tau_1$, $\tau_2$, and A); note that the SHAP values themselves are larger in magnitude compared to k, indicating Q does have an effect on the model output. This agrees with our previous work and understanding that Q primarily governs the cantilever frequency relaxation after the initial perturbation.[3–6] (g)-(i) SHAP value distributions plotted against cantilever resonance frequency ($\omega_0$, kHz) showing how varied resonance frequencies affect the overall model output for each parameter ($\tau_1$, $\tau_2$, and A); from these distributions, we see that the magnitude of SHAP values is relatively small compared to the magnitude of the parameter outputs, suggesting that resonance frequency does not significantly impact the model's prediction.

**Supporting Information Fig 11:** SHAP analysis revealing frequency signal impact on model output for $\tau_1$.

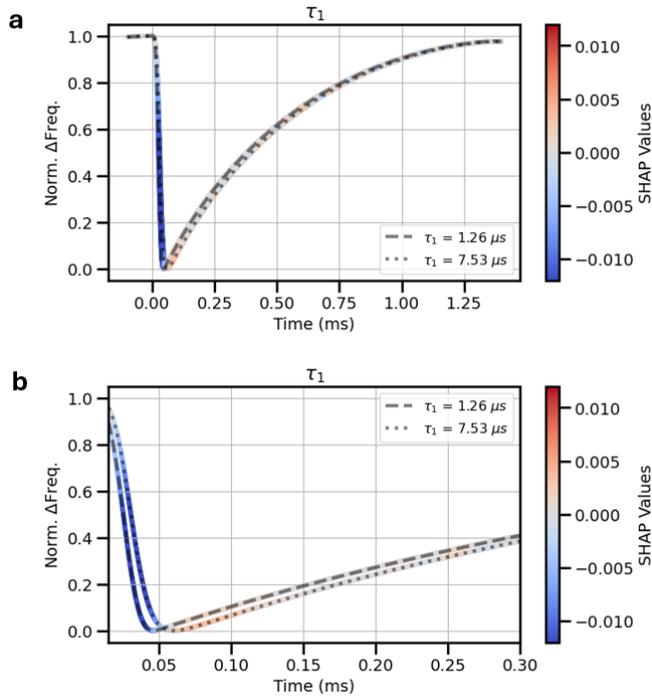

*(a) SHAP analysis revealing initial frequency shift has biggest impact on (highest density of positive/negative SHAP values) on model output corresponding to $\tau_1$. (b) Zoomed in on trigger region to show small density of positive SHAP values that correspond to larger value of $\tau_1$. For these examples, $\tau_2$ was held at 72.53 µs and A was held at 0.9 (a large contribution of $\tau_1$). Examples were simulated using the FFTA code package, freely available at https://github.com/rajgiriUW/ffta.*

**Supporting Information Fig 12** SHAP analysis revealing frequency signal impact on model output for $\tau_2$.

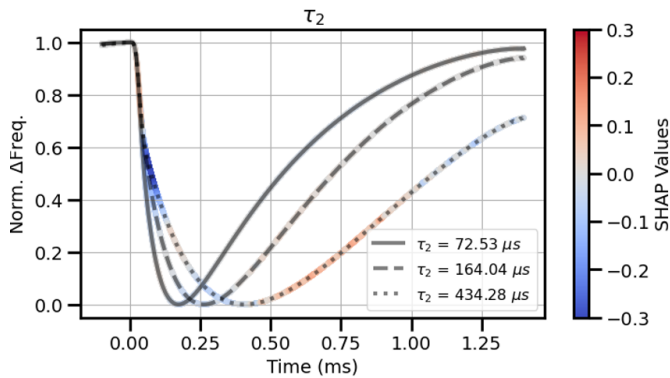

*SHAP analysis revealing initial frequency shift has the biggest impact (highest density of positive/negative SHAP values) on model output corresponding to $\tau_2$.*

**Supporting Information Fig 13:** Individual parameter maps for experimental data shown in main text Figure 5.

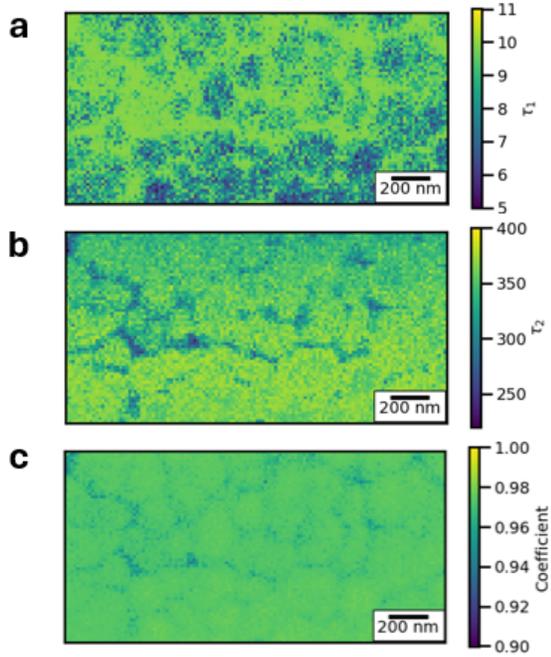

*(a) Multioutput CNN map of extracted $\tau_1$ values (μs) corresponding to trEFM image collected on ITO/Me-4PACz/Cs$_{0.17}$FA$_{0.83}$Pb(I$_{0.85}$Br$_{0.15}$)$_3$ described in main text and shown in Figure 5. (b) Multioutput CNN map of extracted $\tau_2$ values (μs) corresponding to trEFM image described in main text and shown in Figure 6. (c) map of extracted coefficient (A) values corresponding to trEFM image described in main text and shown in Figure 5.*

**Supporting Information Fig 14**: Signal reconstruction $R^2$ map.

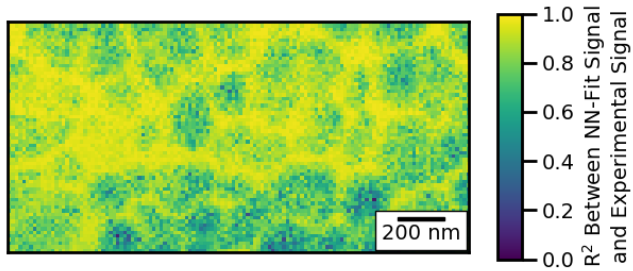

*Map of $R^2$ values comparing experimental data with signals simulated with multioutput CNN parameters, evaluating how well the CNN parameters reconstruct the experimental data.*

**Supporting Information Fig 15**: Single-exponential neural network time constant map and signal reconstruction $R^2$ map for main text Figure 5.

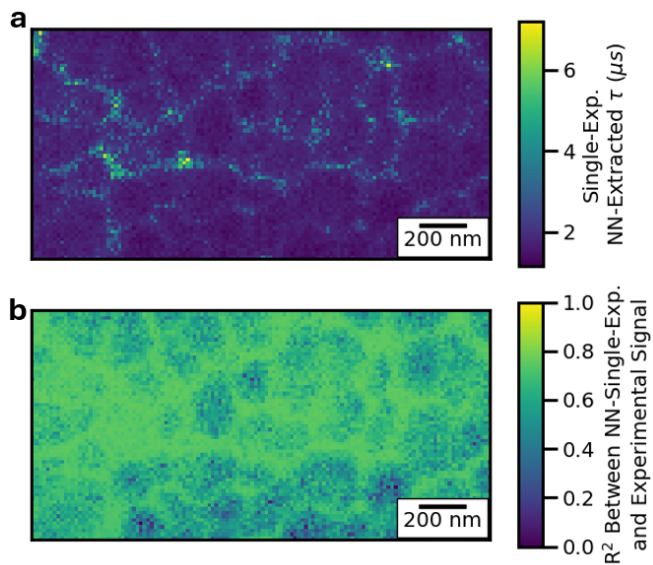

(a) Map of τ values (μs) generated from single-exponential feedforward neural network on experimental trEFM image on ITO/Me-4PACz/$Cs_{0.17}FA_{0.83}Pb(I_{0.85}Br_{0.15})_3$ described in main text and shown in main text Figure 5. (b) Map of $R^2$ values comparing experimental data with signals simulated with single-exponential feedforward neural network[3] parameters, evaluating how well the single-exponential network reconstruct the experimental data.